%Paper: hep-ph/9505227
%From: pati@guinness.ias.edu (Jogesh Pati)
%Date: Wed, 3 May 95 16:21:48 EDT
%Date (revised): Wed, 3 May 95 16:38:46 EDT

%==============================================================
%     TOWARDS A UNIFIED ORIGIN OF FORCES, FAMILIES, AND MASS SCALES
%        Jogesh C. Pati (IAS, Princeton and Univ. of Maryland)
%==============================================================
%  INSTRUCTIONS:
%     Use plain TeX.
%     The accompanying "part 2" contains the Encapsulated PostScript
%          file for Figure 2 which should be printed out SEPARATELY.
%     (Unfortunately, Figure 1 is not in a form that can be submitted
%                         to the bulletin board.)
%==============================================================
% Routine to guarantee that this file is input only once
\catcode`\@=11
\expandafter\ifx\csname @iasmacros\endcsname\relax
	\global\let\@iasmacros=\par
\else	\endinput
\fi
\catcode`\@=12

% Set up font size commands and \baselinestretch command
%  \input IASFONTS

% Some alternative font names
\def\rmb{\seventeenrm}

\def\itb{\twelvebxi}

% Simple spacing commands
\def\singlespace{\baselineskip=\normalbaselineskip}
\def\halfspace{\baselineskip=1.5\normalbaselineskip}
\def\doublespace{\baselineskip=2\normalbaselineskip}

% Macros for references and abstracts

\def\AB{\bigskip\parindent=40pt
        \centerline{\bf ABSTRACT}\medskip\halfspace\narrower}
\def\AE{\bigskip\nonarrower\doublespace}
\def\nonarrower{\advance\leftskip by-\parindent
	\advance\rightskip by-\parindent}

% Useful commands

\def\undertext#1{$\underline{\smash{\hbox{#1}}}$}
\def\boxit#1{\vbox{\hrule\hbox{\vrule\kern3pt
	\vbox{\kern3pt#1\kern3pt}\kern3pt\vrule}\hrule}}

% Special symbols
\def\hence{\leavevmode\hbox{\bf .\raise5.5pt\hbox{.}.} }

\def\dalemb#1#2{{\vbox{\hrule height.#2pt
	\hbox{\vrule width.#2pt height#1pt \kern#1pt \vrule width.#2pt}
	\hrule height.#2pt}}}
\def\gtorder{\mathrel{\raise.3ex\hbox{$>$}\mkern-14mu
             \lower0.6ex\hbox{$\sim$}}}
\def\ltorder{\mathrel{\raise.3ex\hbox{$<$}\mkern-14mu
             \lower0.6ex\hbox{$\sim$}}}

% For twoup output
\newdimen\fullhsize
\newbox\leftcolumn
\def\twoup{\hoffset=-.5in \voffset=-.25in
  \hsize=4.75in \fullhsize=10in \vsize=6.9in
  \def\fullline{\hbox to\fullhsize}
  \let\lr=L
  \output={\if L\lr
        \global\setbox\leftcolumn=\columnbox\global\let\lr=R \advancepageno
      \else \doubleformat \global\let\lr=L\fi
    \ifnum\outputpenalty>-20000 \else\dosupereject\fi}
  \def\doubleformat{\shipout\vbox{
    \fullline{\box\leftcolumn\hfil\columnbox}\advancepageno}}
  \def\columnbox{\leftline{\vbox{\makeheadline\pagebody\makefootline}}}
  \tolerance=1000 }
\catcode`\@=11					% To make protected \def's

%************************************************************
%*
%*		Font set-up
%*
%************************************************************

%************** 5-point fonts *******************************

\font\fiverm=cmr5				% roman
\font\fivemi=cmmi5				% math italic
\font\fivesy=cmsy5				% math symbols
\font\fivebf=cmbx5				% bold face

\skewchar\fivemi='177
\skewchar\fivesy='60

%************** 6-point fonts *******************************

\font\sixrm=cmr6				% roman
\font\sixi=cmmi6				% math italic
\font\sixsy=cmsy6				% math symbols
\font\sixbf=cmbx6				% bold face

\skewchar\sixi='177
\skewchar\sixsy='60

%************** 7-point fonts *******************************

\font\sevenrm=cmr7				% roman
\font\seveni=cmmi7				% math italic
\font\sevensy=cmsy7				% math symbols
\font\sevenit=cmti7				% italic
\font\sevenbf=cmbx7				% bold face

\skewchar\seveni='177
\skewchar\sevensy='60

%************** 8-point fonts *******************************

\font\eightrm=cmr8				% roman
\font\eighti=cmmi8				% math italic
\font\eightsy=cmsy8				% math symbols
\font\eightit=cmti8				% italic
				% slanted
\font\eightbf=cmbx8				% bold face
				% typewriter
				% sans serif

\skewchar\eighti='177
\skewchar\eightsy='60

%************** 9-point fonts *******************************

\font\ninei=cmmi9
\font\ninesy=cmsy9

\skewchar\ninei='177
\skewchar\ninesy='60

%************** 10-point fonts ******************************

\font\tenrm=cmr10				% roman
\font\teni=cmmi10				% math italic
\font\tensy=cmsy10				% math symbols
\font\tenex=cmex10				% math extension
\font\tenit=cmti10				% italic
\font\tensl=cmsl10				% slanted
\font\tenbf=cmbx10				% bold face
\font\tentt=cmtt10				% typewriter
\font\tenss=cmss10				% sans serif
\font\tensc=cmcsc10				% small caps
\font\tenbi=cmmib10				% bold math

\skewchar\teni='177
\skewchar\tenbi='177
\skewchar\tensy='60

\def\tenpoint{\ifmmode\err@badsizechange\else
	\textfont0=\tenrm \scriptfont0=\sevenrm \scriptscriptfont0=\fiverm
	\textfont1=\teni  \scriptfont1=\seveni  \scriptscriptfont1=\fivemi
	\textfont2=\tensy \scriptfont2=\sevensy \scriptscriptfont2=\fivesy
	\textfont3=\tenex \scriptfont3=\tenex   \scriptscriptfont3=\tenex
	\textfont4=\tenit \scriptfont4=\sevenit \scriptscriptfont4=\sevenit
	\textfont5=\tensl
	\textfont6=\tenbf \scriptfont6=\sevenbf \scriptscriptfont6=\fivebf
	\textfont7=\tentt
	\textfont8=\tenbi \scriptfont8=\seveni  \scriptscriptfont8=\fivemi
	\def\rm{\tenrm\fam=0 }%
	\def\it{\tenit\fam=4 }%
	\def\sl{\tensl\fam=5 }%
	\def\bf{\tenbf\fam=6 }%
	\def\tt{\tentt\fam=7 }%
	\def\ss{\tenss}%
	\def\sc{\tensc}%
	\def\bmit{\fam=8 }%
	\rm\setparameters\setbaselines\fi}

%************** 12-point fonts ******************************

\font\twelverm=cmr12				% roman
\font\twelvei=cmmi12				% math italic
\font\twelvesy=cmsy10	scaled\magstep1		% math symbols
\font\twelveex=cmex10	scaled\magstep1		% math extension
\font\twelveit=cmti12				% italic
\font\twelvesl=cmsl12				% slanted
\font\twelvebf=cmbx12				% bold face
\font\twelvett=cmtt12				% typewriter
\font\twelvess=cmss12				% sans serif
\font\twelvesc=cmcsc10	scaled\magstep1		% small caps
\font\twelvebi=cmmib10	scaled\magstep1		% bold math
\font\twelvebxi=cmbxti10 scaled\magstep1	% bold italic

\skewchar\twelvei='177
\skewchar\twelvebi='177
\skewchar\twelvesy='60

\def\twelvepoint{\ifmmode\err@badsizechange\else
	\textfont0=\twelverm \scriptfont0=\eightrm \scriptscriptfont0=\sixrm
	\textfont1=\twelvei  \scriptfont1=\eighti  \scriptscriptfont1=\sixi
	\textfont2=\twelvesy \scriptfont2=\eightsy \scriptscriptfont2=\sixsy
	\textfont3=\twelveex \scriptfont3=\tenex   \scriptscriptfont3=\tenex
	\textfont4=\twelveit \scriptfont4=\eightit \scriptscriptfont4=\sevenit
	\textfont5=\twelvesl
	\textfont6=\twelvebf \scriptfont6=\eightbf \scriptscriptfont6=\sixbf
	\textfont7=\twelvett
	\textfont8=\twelvebi \scriptfont8=\eighti  \scriptscriptfont8=\sixi
	\def\rm{\twelverm\fam=0 }%
	\def\it{\twelveit\fam=4 }%
	\def\sl{\twelvesl\fam=5 }%
	\def\bf{\twelvebf\fam=6 }%
	\def\tt{\twelvett\fam=7 }%
	\def\ss{\twelvess}%
	\def\sc{\twelvesc}%
	\def\bmit{\fam=8 }%
	\rm\setparameters\setbaselines\fi}

%************** 14-point fonts ******************************

\font\fourteenrm=cmr10	scaled\magstep2		% roman -- talaris=cmr10 \step2
\font\fourteeni=cmmi10	scaled\magstep2		% math italic
\font\fourteensy=cmsy10	scaled\magstep2		% math symbols
\font\fourteenex=cmex10	scaled\magstep2		% math extension
\font\fourteenit=cmti10	scaled\magstep2		% italic
\font\fourteensl=cmsl10	scaled\magstep2		% slanted
\font\fourteenbf=cmbx10	scaled\magstep2		% bold face
\font\fourteentt=cmtt10	scaled\magstep2		% typewriter
\font\fourteenss=cmss10	scaled\magstep2		% sans serif
\font\fourteensc=cmcsc10 scaled\magstep2	% small caps
\font\fourteenbi=cmmib10 scaled\magstep2	% bold math

\skewchar\fourteeni='177
\skewchar\fourteenbi='177
\skewchar\fourteensy='60

\def\fourteenpoint{\ifmmode\err@badsizechange\else
	\textfont0=\fourteenrm \scriptfont0=\tenrm \scriptscriptfont0=\sevenrm
	\textfont1=\fourteeni  \scriptfont1=\teni  \scriptscriptfont1=\seveni
	\textfont2=\fourteensy \scriptfont2=\tensy \scriptscriptfont2=\sevensy
	\textfont3=\fourteenex \scriptfont3=\tenex \scriptscriptfont3=\tenex
	\textfont4=\fourteenit \scriptfont4=\tenit \scriptscriptfont4=\sevenit
	\textfont5=\fourteensl
	\textfont6=\fourteenbf \scriptfont6=\tenbf \scriptscriptfont6=\sevenbf
	\textfont7=\fourteentt
	\textfont8=\fourteenbi \scriptfont8=\tenbi \scriptscriptfont8=\seveni
	\def\rm{\fourteenrm\fam=0 }%
	\def\it{\fourteenit\fam=4 }%
	\def\sl{\fourteensl\fam=5 }%
	\def\bf{\fourteenbf\fam=6 }%
	\def\tt{\fourteentt\fam=7}%
	\def\ss{\fourteenss}%
	\def\sc{\fourteensc}%
	\def\bmit{\fam=8 }%
	\rm\setparameters\setbaselines\fi}

%************** Miscellaneous big fonts *********************

\font\seventeenrm=cmr10 scaled\magstep3		% roman
		% bold face

%************************************************************
%*
%*		Parameter initialization
%*
%************************************************************

\newdimen\rp@
\newcount\@basestretchnum
\newskip\@baseskip
\newskip\headskip
\newskip\footskip

% Routine to set page parameters

\def\setparameters{\rp@=.1em
	\headskip=24\rp@
	\footskip=\headskip
	\delimitershortfall=5\rp@
	\nulldelimiterspace=1.2\rp@
	\scriptspace=0.5\rp@
	\abovedisplayskip=10\rp@ plus3\rp@ minus5\rp@
	\belowdisplayskip=10\rp@ plus3\rp@ minus5\rp@
	\abovedisplayshortskip=5\rp@ plus2\rp@ minus4\rp@
	\belowdisplayshortskip=10\rp@ plus3\rp@ minus5\rp@
	\normallineskip=\rp@
	\lineskip=\normallineskip
	\normallineskiplimit=0pt
	\lineskiplimit=\normallineskiplimit
	\jot=3\rp@
	\setbox0=\hbox{\the\textfont3 B}\p@renwd=\wd0
	\skip\footins=12\rp@ plus3\rp@ minus3\rp@
	\skip\topins=0pt plus0pt minus0pt}

% Special routine to scale \baselineskip

\def\setbaselines{\maxdepth=4\rp@\baselinestretch=\@basestretchnum}

% The \baselinestretch command

\def\baselinestretch{\afterassignment\@basestretch\@basestretchnum}
\def\@basestretch{%
	\@baseskip=12\rp@ \divide\@baseskip by1000
	\normalbaselineskip=\@basestretchnum\@baseskip
	\baselineskip=\normalbaselineskip
	\bigskipamount=\the\baselineskip
		plus.25\baselineskip minus.25\baselineskip
	\medskipamount=.5\baselineskip
		plus.125\baselineskip minus.125\baselineskip
	\smallskipamount=.25\baselineskip
		plus.0625\baselineskip minus.0625\baselineskip
	\setbox\strutbox=\hbox{\vrule height.708\baselineskip
		depth.292\baselineskip width0pt }}

%************************************************************
%*
%*		Modifications to PLAIN.TEX
%*
%************************************************************

% Modifications to PLAIN routines to handle scaling of page parameters

\def\makeheadline{\vbox to0pt{\baselinestretch=1000
	\vskip-\headskip \vskip1.5pt
	\line{\vbox to\ht\strutbox{}\the\headline}\vss}\nointerlineskip}

\def\makefootline{\baselineskip=\footskip\line{\the\footline}}

\def\big#1{{\hbox{$\left#1\vbox to8.5\rp@ {}\right.\n@space$}}}
\def\Big#1{{\hbox{$\left#1\vbox to11.5\rp@ {}\right.\n@space$}}}
\def\bigg#1{{\hbox{$\left#1\vbox to14.5\rp@ {}\right.\n@space$}}}
\def\Bigg#1{{\hbox{$\left#1\vbox to17.5\rp@ {}\right.\n@space$}}}

% Modifications to PLAIN to handle bold math

\mathchardef\alpha="710B
\mathchardef\beta="710C
\mathchardef\gamma="710D
\mathchardef\delta="710E
\mathchardef\epsilon="710F
\mathchardef\zeta="7110
\mathchardef\eta="7111
\mathchardef\theta="7112
\mathchardef\iota="7113
\mathchardef\kappa="7114
\mathchardef\lambda="7115
\mathchardef\mu="7116
\mathchardef\nu="7117
\mathchardef\xi="7118
\mathchardef\pi="7119
\mathchardef\rho="711A
\mathchardef\sigma="711B
\mathchardef\tau="711C
\mathchardef\upsilon="711D
\mathchardef\phi="711E
\mathchardef\chi="711F
\mathchardef\psi="7120
\mathchardef\omega="7121
\mathchardef\varepsilon="7122
\mathchardef\vartheta="7123
\mathchardef\varpi="7124
\mathchardef\varrho="7125
\mathchardef\varsigma="7126
\mathchardef\varphi="7127
\mathchardef\imath="717B
\mathchardef\jmath="717C
\mathchardef\ell="7160
\mathchardef\wp="717D
\mathchardef\partial="7140
\mathchardef\flat="715B
\mathchardef\natural="715C
\mathchardef\sharp="715D

%************************************************************
%*
%*		Initialization
%*
%************************************************************

\def\err@badsizechange{%
	\immediate\write16{--> Size change not allowed in math mode, ignored}}

\baselinestretch=1000
\tenpoint

\catcode`\@=12					% Restore @ sign

\twelvepoint
\hsize=6.5in
\baselineskip=10pt
\overfullrule=0pt
\rightline{IASSNS-HEP-95/11}
\rightline{hep-ph/9505227}
\rightline{April 1995}
\bigskip
\centerline{\rmb TOWARDS A UNIFIED ORIGIN OF FORCES,
FAMILIES}
\smallskip
\centerline{{\rmb  AND MASS SCALES}\footnote{$^\dagger$}
{Talk presented at the ``1994 Int'l. Workshop on B-Physics,'' Nagoya,
Japan (Oct. 26-29, 1994), at the Int'l. Symposium on ``Physics at the
Planck Scale,'' Puri, India, Dec. 10-20, 1994, and at the DAE Symposium
on High Energy Physics, Shantiniketan (India), Dec. 28-Jan. 2, 1995.}}

\bigskip
\centerline{\itb Jogesh C. Pati}
\centerline{\bf Institute for Advanced Study}
\centerline{\bf Princeton, NJ 08540 \ USA}
\centerline{\bf and}
\centerline{\bf Department of Physics,  University of Maryland}
\centerline{\bf College Park, MD  \ 20742 \ USA}

{\AB
{\baselineskip=8pt
In spite of the success of the conventional approach to grand
unification as regards the meeting of the gauge coupling constants, it
is expressed that the associated arbitrariness in the choice of the
Higgs-sector parameters goes against the very grain of unification.
Owing to this arbitrariness, the Higgs-exchange force is in fact still
ununified.  The superstring theories, following the assumption of
elementary quarks, have not yet yielded the {\it right package} of
Higgs-sector parameters, through any one of their plethora of
solutions -- a task that {\it a priori} seems a heavy burden.

A case is therefore made for an alternative approach to unification that
is based on a {\it purely gauge origin of the fundamental forces},  and is
thus devoid of the Higgs-sector altogether.  This approach seems to call
for the ideas of local supersymmetry and preons.  The associated
spectrum and forces may well have their origin within a superstring
theory, which would, however, be relieved (in this case) from the burden
of yielding the ``right package'' of Higgs-sector parameters, because
the system has no Higgs sector.  The advantage of the marriage of the
ideas of local supersymmetry and preons, subject to two broad dynamical
assumptions which are specified, are noted.  These include true economy
and viability as well as an understanding of the origins of (a)
family-replication, (b) inter-family mass-hierarchy, and (c) diverse
mass-scales which span from $M_{Planck}$ to $m_W \sim m_t$ to $m_e$ to
$m_\nu$.  In short, the approach seems capable of providing {\it a
unified origin of the forces, the families and the mass-scales}.  In the
process, the preonic approach provides the scope for synthesizing a rich
variety of phenomena all of which could arise dynamically through one
and the same tool -- the SUSY metacolor force coupled with gravity -- at
the scale of $10^{11}GeV$.  The phenomena include:  (i) spontaneous
violations of parity, CP, B-L and Peccei-Quinn symmetry,
(ii) origin of heavy Majorana mass for $\nu_R$, (iii) SUSY breaking,
(iv) origins of
even $m_W,~m_q$ and $m_\ell$, as well as, (v) inflation and
lepto/baryo-genesis.

Some intriguing experimental consequences of the new approach which
could show at LEPI, LEPII and Tevatron and a {\it crucial prediction}
which can be probed at the LHC and NLC are presented.
\AE
}}
\vfill\eject
\pageno=2
\singlespace

\item{\rmb 1.}
{\rmb Going Beyond the Standard Model}

The standard model of particle physics (SM) has brought a good deal of
synthesis in our understanding of the basic forces of nature, especially
in comparison to its predecessors, and has turned out to be brilliantly
successful in terms of its agreement with experiments.  Yet, as
recognized for sometime [1], it falls short as a fundamental theory
because it introduces some 19 parameters.  And it
does not explain (i) family replication;
(ii) the coexistence of the two kinds of matter:  quarks {\it
and} leptons; (iii) the coexistence of the electroweak {\it and} the
QCD forces with their hierarchical strengths $g_1 \ll g_2 \ll
g_3$, as observed at low energies; (iv) quantization of electric charge;
(v) inter and intrafamily mass-hierarchies - {\it i.e.},
$m_{u, d, e} \ll m_{c,s,\mu}\ll m_{t,b,\tau}$ and $m_b\ll m_t$, etc. -
reflected by ratios such as $(m_u/m_t) \sim 10^{-4},~(m_c/m_t)\sim
10^{-2}$ and $(m_b/m_t)\sim {1\over 35}$; and (vi) the origin of diverse
mass scales that span over more than 27 orders of magnitude from
$M_{Planck}$ to $m_W$ to $m_e$ to $m_\nu$, whose ratios involve
very small numbers such as$(m_W/M_{Pl})\sim
10^{-17},~(m_e/M_{Pl})\sim 10^{-22}$ and $(m_\nu/M_{Pl})< 10^{-27}$.
There are in addition the two most basic questions:  (vii) how does
gravity fit into the whole scheme, especially in the context of a good
quantum theory?, and (viii) why is the cosmological constant so small or
zero?

These issues constitute at present some of the major puzzles of particle
physics and provide motivations for contemplating new
physics beyond the standard model which should shed light on them.
The ideas which have been proposed and which do show promise to
resolve at least some of these puzzles include:
\medskip
(1)~~{\bf Grand Unification}: ~~ The hypothesis of grand
unification, which proposes an underlying unity of the fundamental
particles and their forces [1,2,3],
appears attractive
because it explains at once (i) the quantization of electric charge, (ii) the
existence of quarks {\it and} leptons with $Q_e=-Q_p$, and (iii) the
existence of the strong, the electromagnetic and the weak forces with
$g_3\gg g_2\gg g_1$ at low energies, but $g_3=g_2=g_1$ at high energies.
These are among the puzzles listed above and grand unification
resolves all three.
{\it Therefore I believe that the central concept of
grand unification is, very likely, a step in the right direction.}  By
itself, it does not address, however, the remaining
puzzles listed above, including the issues of family replication and
origin of mass-hierarchies.
\medskip
(2)~~{\bf Supersymmetry}:~~ This is the symmetry that
relates fermions to bosons[4].  As a local
symmetry, it is attractive because it implies the existence of gravity.
It has
the additional virtue that it
helps maintain a large hierarchy in mass-ratios such as
$(m_{\phi}/M_U) \sim 10^{-14}$ and $(m_{\phi}/M_{p\ell}) \sim 10^{-17}$,
without
the need for fine tuning, provided, however, such ratios are put in by
hand.
Thus it provides a technical resolution of the gauge hierarchy problem,
{\it but
by itself does not explain the origin of the large hierarchies}.
\medskip
(3)~~{\bf Compositeness}:~~  Here there are {\it two
distinct suggestions}:
\smallskip
(a)~~\undertext{Technicolor}:  The idea of technicolor [5]
proposes that the Higgs bosons are composite but quarks and leptons are
still elementary.  Despite the attractive feature of dynamical symmetry
breaking which eliminates elementary Higgs bosons and thereby the arbitrary
parameters which go with them, this idea
is excluded, at least in its simpler versions, owing to conflicts with
flavor-changing neutral current processes and oblique electroweak
corrections.  The so-called walking technicolor models may be arranged
to avoid some of these conflicts at the expense, however, of excessive
proliferation in elementary constituents.  Furthermore, as a generic
feature, none of these models seem capable of addressing any of the
basic issues listed above, including those of family replication and
fermion mass-hierarchies.  Nor do they go well with the
hypothesis of a unity of the basic forces.
\smallskip
(b)~~\undertext{ Preons}:  By contrast, the idea of preonic
compositeness which proposes that not just the Higgs bosons but also the
quarks and the leptons are composites of a {\it common} set of
constituents called ``preons'' seems much more promising.  Utilizing
supersymmetry to its advantage, the preonic approach has evolved over
the last few years to acquire a form [6,7,8,9] which is (a) far more
economical in field-content and especially in parameters than either the
technicolor or the conventional grand unification models, and, (b)
is viable.  Most important, utilizing primarily the symmetries of the
theory (rather than detailed dynamics) and the peculiarities of SUSY QCD
as regards forbiddeness of SUSY-breaking, in the absence of gravity,
the preonic
approach provides simple explanations for the desired protection of
composite quark-lepton masses and at the same time for the origins of
family-replication, inter-family mass-hierarchy and diverse mass scales.
It also provides several testable predictions.  In this sense, though
still unconventional, the preonic approach shows promise in being able
to address certain fundamental issues.  I will return to it shortly.
\medskip
(4)~~{\bf Superstrings}: ~~Last but not least, the idea of
superstrings [10] proposes that the elementary entities are not truly
pointlike but are extended stringlike objects with sizes $\sim
(M_{Planck})^{-1}
\sim 10^{-33}$ cm.  This idea appears to be most promising in providing a
unified
theory of all the forces of nature including gravity and yielding a
well-behaved
quantum theory of gravity.  In principle, assuming that quarks, leptons
and Higgs bosons are elementary, a suitable superstring theory
could also account for the origin of the three families and the Higgs
bosons at the string unification scale, as well as explain all the
parameters of the SM. But in practice, this has not happened as yet.
Some general stumbling blocks of string theories are associated with
the problems of (i) a choice of the ground state (the vacuum)
from among the many solutions and (ii) understanding supersymmetry breaking.

The ideas listed above are, of course, not mutually exclusive.  In fact the
superstring theories already comprise local supersymmetry and the central
idea of grand unification.  It remains to be seen, however, whether
they give rise, in accord with the standard belief, to elementary
quarks and leptons, or alternatively to a set of substructure fields --
the preons.  In the following, I first
recall the status of conventional grand unification,
and then provide
a perspective as well as motivations for an alternative approach to
grand unification, based on the idea of preons.
\medskip
\noindent
{\rmb 2.}~~{\rmb Grand Unification in the Conventional Approach}
\smallskip
\noindent
{\bf (A)}~~{\bf The SU(5) vs. SU(4)-Color Routes:  A Distinction
through Neutrinos}

By ``Conventional approach'' to grand unification I mean the one in
which quarks and leptons -- and traditionally the Higgs bosons as well
-- are assumed to be elementary [1,2,3].  Within this approach, there
are two distinct routes to higher unification:  (i) the SU(4)-color
route [1] and (ii) SU(5)[2].
Insisting on a compelling reason
for charge -- quantization, the former naturally introduces the
left-right symmetric gauge structure $G_{224}=SU(2)_L \times
SU(2)_R\times SU(4)_{L+R}^C$ [1], which in turn may be embedded in
anomaly-free simple groups like SO(10) or $E_6$ [11].

While the two routes share some important common features, such as
quark-lepton unification and unification of gauge coupling constants
(see discussions below), there are two major
conceptual and practical distinctions between them.
SU(5), {\it viewed as
fundamental}, (rather than part of SO(10), (i) violates parity
explicitly and (ii) does not have a
compelling reason for the existence of $\nu_R$'s and therefore for the
$\nu_L$'s to be massive.  Even allowing for higher dimensional operators
induced by Planck scale physics of the type
$\nu_L\nu_L\langle\phi\rangle\langle\phi\rangle/M$ one obtains
$m(\nu_L)\sim 10^{-5}-10^{-4} eV$ for $M\sim M_{Pl}\sim 10^{19}-10^{18}
GeV$, which is too small for neutrinos to serve as HDM and (perhaps) a
bit too small even for them to be relevant to the MSW explanation of the
solar neutrino puzzle.

By contrast, the SU(4)-color route in the context of the core symmetry
$G_{224}$ (and therefore higher symmetries like SO(10) or $E_6$ as
well) (i) proposes parity to be exact at high energies which is
violated only spontaneously [12] and (ii) {\it requires} the existence
of $\nu_R$'s and therefore neutrinos to have at least Dirac masses.
Together with heavy Majorana masses for $\nu_R$'s of order
$10^{11}-10^{12} GeV$, which can arise spontaneously together with
$SU(2)_R$-breaking in a variety of ways [13],
one obtains the light
see-saw neutrino masses $ m(\nu_L^i)\simeq
(m(\nu^i)_{Dirac}^2/M_{\nu_R})$.  This yields the mass-pattern:
$m(\nu_L^e)\simeq 10^{-8}~eV,~m(\nu_L^\mu)\simeq (10^{-3}-10^{-1})~eV$
and $m(\nu_L^\tau)\simeq (1~{\rm to}~ 30)~eV$, which goes well both with
the MSW solution for the solar neutrino puzzle and with the candidacy of
$\nu_\tau$ for HDM.  Thus, if neutrino masses settle at the presently
indicated values, especially if at least one of them has a mass of
order one to few
electron volts, as suggested by the need for HDM, one would
have a strong indication in favor of the
SU(4)-color route and left-right symmetry, as opposed to the
SU(5)-route.

The other distinguishing feature of SU(4)-color is that it
gauges B-L as a local
symmetry.  Following the old arguments of Lee and Yang, based on
E\"otvos-type experiments, it follows that the massless gauge particle
coupled to B-L must acquire mass spontaneously and thereby B-L must
break spontaneously.  Such B-L violations at an intermediate scale $(\sim
10^{11}~GeV$) may well be necessary to implement baryogenesis, since
electroweak effects erase B-L conserving baryon excess, generated at
higher temperatures.  {\it Because of these desirable
features, I believe that the symmetry group $G_{224}$ is likely to be
part of a fundamental theory}.  It turns out that even the preonic
approach to unification (to be presented here) relies heavily on the
symmetry group $G_{224}$ and thus enjoys the same advantages, as regards
neutrino masses and B-L violation.  I now turn to the two central
features of grand unification.
\medskip
\noindent
{\bf (B)}~~
{\bf Meeting of the Coupling Constants and Proton Decay}

It has been known for sometime that the dedicated proton decay
searches at the IMB and the Kamiokande detectors [14], and more
recently the precision measurements of the SM coupling constants (in
particular ${\rm sin}^2 \hat\theta_W$) at LEP [15] put severe constraints on
grand unification models without supersymmetry.  Owing to such
constraints, the non-SUSY minimal SU(5) and, for similar reasons, the
one-step breaking non-SUSY SO(10)-model, as well, are now excluded
beyond a shadow of doubt.

But the idea of the union of the coupling constants $g_1, g_2,$ and
$g_3$ can well materialize in accord with the LEP data, if one invokes
supersymmetry [16,17,18] into minimal SU(5) or SO(10).  See Fig. 1, which
shows the {\it impressive meeting} of the three coupling constants of
the minimal supersymmetric standard model (MSSM) with an assumed
SUSY-threshold around 1 TeV.  Such a model can, of course, be embedded
within a minimal SUSY SU(5) or SO(10) model, which would provide the
rationale for the meeting of the coupling constants at a scale $M_U
\approx 2 \times 10^{16}$ GeV.

The fact that the coupling constants meet in the context of these models
is reflected by the excellent agreement of their predicted value of
$[{\rm sin}^2 \hat\theta_W (m_z)_{theory}=\cdot 2325\pm \cdot 005$ (using
$\alpha_s(m_z)=\cdot 12\pm \cdot 01)$ with that determined at LEP:
$[{\rm sin}^2\hat\theta_W(m_z)]_{expt.}=\cdot 2316\pm\cdot 0003$.
In SUSY SU(5) or SO(10), dimension 5 operators do in general pose
problems for proton decay.  But the relevant parameters of the
SUSY-space can be arranged
to avoid conflict with experiments [19].  The SUSY-extensions of SU(5) or
SO(10) typically lead to prominent strange particle decay modes, e.g.,
$p\to \bar\nu K^+$ and $n\to \bar\nu K^0$, while a 2-step breaking of
SO(10) via the intermediate symmetry $G_{224}$ can also lead to
prominent $\Delta (B-L)=-2$ decay modes of the nucleon via Higgs
exchanges such as $p\to e^-\pi^+\pi^+$ and $n\to e^-\pi^+$ and even
$n\to e^-e^+\nu_e$, etc. in addition to the canonical $e^+\pi^0$-mode
[20].

It is encouraging that the super-Kamiokande (to be completed in April
1996) is expected to be sensitive to the $e^+\pi^0$ mode up to partial
lifetimes of few $\times 10^{34}$ years, to the $\bar\nu K^+$ and
$\bar\nu K^0$ modes with partial lifetimes $\leq 10^{34}$ years and to
the non-canonical $n\to e^-e^+\nu_e$ and $p\to e^-\pi^+\pi^+$ modes with
partial lifetimes $< 10^{33}$ years.  Thus the super-Kamiokande,
together with other forthcoming
facilities, in particular, ICARUS,
provide a {\it big
ray of hope} that first of all one will be able to probe much deeper
into neutrino physics in the near future and second proton-decay may
even be discovered within the twentieth century.
\medskip
\noindent
{\bf (C)}~~{\bf Questioning the Conventional Approach}

Focusing attention on the meeting of the coupling constants (Fig. 1),
the question arises:  To what extent does this meeting  reflect the
``truth'' or
is it somehow deceptive?  There are two reasons why such a question is
in order.
\smallskip
(1)~~ First, the unity of forces reflected by the meeting of the
coupling constants in SUSY SU(5) or SO(10) is truly incomplete,
because it comprises only the gauge forces, but not the
Higgs-exchange forces.  {\it The latter are still governed by many arbitrary
parameters -- i.e., the masses, the quartic and the Yukawa couplings of
the Higgs bosons  -- and are thus ununified.}  Such arbitrariness goes
against the central spirit of grand unification and has been the main
reason in my mind since the 1970's (barring an important caveat due to
the growth of superstring theories in the 1980's, see below) to consider
seriously the possibility that the Higgses as well as the quarks and the
leptons are composite.  Furthermore, neither SUSY
SU(5) nor SUSY SO(10),
by itself, has the scope of explaining the origins of (a) the three
families, (b) inter- and intra-family mass-splittings and (c) the
hierarchical mass-scales:  from $M_{Planck}$ to $m_\nu$.
\smallskip
(2)~~ The second reason for questioning the conventional approach is
this:  one might have hoped that one of the two schemes -- i.e.,
the minimal SUSY SU(5) or the SUSY SO(10)-model, or a broken ``grand
unified'' symmetry with relations between its gauge couplings near the
string scale, would emerge from one of
the solutions of the superstring theories [10,21], which would yield not
only the desired
spectrum of quarks, leptons and Higgs bosons but also just the right
parameters for the Higgs masses as well as their quartic and Yukawa
couplings.  While it seems highly nontrivial that so
many widely varying parameters should come out in just the right way
simply from topological and other constraints of string theories,
it would of course be most
remarkable if that did happen.  {\it But so far it has not}.  There are in
fact a very large number of classically allowed degenerate 4D solutions of the
superstring theories (Calabi-Yau, orbifold and free fermionic, etc.),
although one is not yet able
to choose between them.  Notwithstanding this general difficulty of a
choice, it
is interesting that there are at least some
three-family solutions.  However, not
a single one of these has yielded {\it either} a SUSY SU(5) or an
SO(10)-symmetry, {\it or} a broken ``grand unified'' symmetry involving
direct product of groups, with the desired spectrum
{\it and} parameters, so as to explain the bizarre pattern of fermion masses
and mixings of the three families [27].  Note that for a string theory
to yield elementary quarks, leptons and Higgs bosons, either the {\it
entire package} of calculable parameters, which describe the masses of
all the fermions and their mixings (subject to perturbative
renormalization), should come out just right, or else the corresponding
solution must be discarded.  This no doubt is a {\it heavy burden}.
For the case of the broken grand unified models, there is
the additional difficulty that the grand unification scale of $2\times
10^{16}~GeV$ obtained from low-energy extrapolation does not match
the string unification scale of about $4\times
10^{17}~GeV$ [23].

Thus, even if a certain superstring theory is the right starting point,
and I believe it is, it is not at all clear, especially in view of the
difficulties mentioned above, that it makes contact with the low-energy
world by yielding elementary quarks, leptons and Higgs bosons.  In this
sense, it seems prudent to keep open the possibility that the meeting of
the coupling constants in the context of conventional grand unification,
which after all corresponds to predicting just one number -- i.e.,
$sin^2\theta_W$ -- correctly, may be fortuitous.  Such a meeting should
at least be
viewed with caution as regards inferring the extent to which it
reflects the ``truth'' because there are in fact alternative ways by
which such a meeting can occur (see discussions below).
\smallskip
\noindent
{\bf (D)}~~{\bf Motivations for the Preonic Alternative}

This brings me to consider an alternative approach to unification based
on the ideas of preons and local supersymmetry [6,7,8,9].  Although the
general idea of preons is old, the particular approach which I am about
to present
has evolved in the last few years.  It is still unconventional, yet I
believe, it is promising.  Its
lagrangian consists of only the minimal gauge interactions.
{\it It is devoid altogether of the Higgs sector} {\it since its
superpotential is zero owing to gauge and non-anomalous R-symmetry},
and therefore it is free from all the arbitrary
mass, quartic and Yukawa coupling parameters which conventionally go
with this sector.  This brings {\it real economy}.  In fact, the preon
model possesses only three (or four) parameters which correspond to its
gauge couplings (see below) and even these few would merge into one near
the Planck scale if there is an underlying unity of forces
as we envisage [24].  By contrast, the standard model has 19
and conventional SUSY grand unification models have over 15 parameters.
As mentioned in the introductory chapter, in addition to economy, the main
motivations for pursuing the preonic approach are that
it provides simple explanations for (a) the
protection of the masses of the composite quarks and leptons,
(b) family replication,
(c) inter-family mass-hierarchy $(m_{u,d,e} \ll m_{c,s,\mu}\ll
m_{t,b,\tau}$) and (d) diverse mass-scales.  It still preserves
the central spirit of grand unification [24] and it is viable with
respect to observed processes including flavor-changing neutral current
processes (see remarks later) and oblique EW corrections.

Certain novel features in the dynamics of a class of SUSY QCD theories,
in particular the forbidding of SUSY-breaking in the absence of
gravity, and symmetries of the underlying preonic theory, play crucial
roles in achieving these desired results (a)--(d).  Furthermore, the approach
provides some {\it crucial tests}, which can be performed at the LHC and
possibly LEP 200, by which it can be vindicated if it is right or
excluded if it is wrong.  To present some of these results, I first
need to recall a few salient features of the associated model.
\bigskip
\noindent
{\rmb 3.}~~{\rmb The Preon Model:~~Challenges and Advantages}
\smallskip
\noindent
{\bf (A)~~Lagrangian:} ~~
The model [6] is defined through an effective lagrangian, just below the
Planck scale, possessing $N$ = 1 local supersymmetry and a gauge symmetry
of
the form $G_M \times G_{fc}$.  Here the symmetry $G_M = SU(N)_M$ (or
$SO(N)_M$)
generates an asymptotically free metacolor gauge force which binds preons
and
$G_{fc}$ denotes the flavor-color gauge symmetry, which is assumed to be
either
$G_{224} = SU(2)_L \times SU(2)_R \times SU(4)^C$ [1], or a subgroup of
$G_{224}$
containing $SU(2)_L \times U(1) \times SU(3)^C$.
In what follows, to be specific, we assume $G_M=SU(N)_M$, although
$SO(N)_M$ is a feasible choice.  The gauge symmetry $G_M
\times
G_{fc}$ operates on a set of preonic constituents consisting of six
positive and
six negative {\it massless} chiral superfields $\Phi^a_{\pm} =
(\varphi^a_{L,R},
\psi^a_{L,R})$.  Each of these transforms as the fundamental
representation ${\underline N}$ of $G_M = SU(N)_M$.  The index $a$ runs
over six
values: \ $(x,y); (r,y,b,l)$, where $(x,y)$ denote the two basic
flavor-attributes $(u,d)$ and $(r,y,b,l)$ the four basic color-attributes
of a
quark-lepton family [1].  Thus $\Phi_+^{x,y}$ and $\Phi_-^{x,y}$
transform
as doublets of $SU(2)_L$ and $SU(2)_R$ respectively, while both
$\Phi_+^{r,y,b,l}$ and $\Phi_-^{r,y,b,l}$ transform as quartets of
$SU(4)^C$.
The effective lagrangian of this model turns out to possess only gauge and
gravitational interactions.  As a result, the model involves {\it at most
only
three or four parameters} (see below) corresponding to the coupling
constants of
the gauge symmetry $G_M \times G_{fc}$.
For a number
of reasons, including unity of forces, it turns out that $N$ of $SU(N)_M$
should be between 6 and 4; thus $N_{flavors}=6=N$ or $N+1$ or $N+2$.
Note, since ``a'' runs over six values, the global symmetry of the
metacolor force is $G=SU(6)_L\times SU(6)_R\times U(1)_V\times U(1)_X$
where $U(1)_V$ is the preon number and $U(1)_X$ is the non-anomalous $R$
symmetry [25].  In the presence of flavor-color interactions the
$SU(6)_L\times SU(6)_R$ symmetry is, of course, approximate.  Only its
subgroup $G_{224}=SU(2)_L\times SU(2)_R\times SU(4)_{L+R}^c$ (or a
subgroup of $G_{224}$ containing the SM symmetry $G_{213}$), which is
gauged, together with two global $U(1)$'s (up to QCD and EW anomalies),
is exact.

Such a model has not yet been derived from a superstring theory, although
there
does not appear to be any bar, in principle in this regard especially in
the
context of four-dimensional fermionic constructions [21].  Even without
such a
derivation, if one introduces the economical preon-picture mentioned above
through an effective lagrangian just below the Planck scale and assumes $N$
= 1
local supersymmetry, one seems to be able to derive a number of
advantages, subject to a few broad dynamical
assumptions, which are stated below.  These assumptions seem to me at
least not
implausible and,
more important, they are
{\it not} trivially related to the advantages.
\smallskip
\noindent
{\bf (B)~~Dynamical Assumptions and Challenges}:~~
The {\it two main assumptions} of the model are:
(1) The asymptotically free SUSY metacolor force with
massless preons and (at least) one paired set of values of $N_f$ and
$N$, mentioned above, confines [26].  (2) As the metacolor (MC) force
becomes strong at a scale $\Lambda_M$, which turns out to be of order
$10^{11}~GeV$ (see discussions later), it makes a few metacolor
singlet composites, which
include quarks and leptons (their replication and the protection of
their masses will be justified on independent grounds).
Furthermore, it makes a few MC singlet preonic condensates which break
the (approximate) global symmetry $G$ as well as its gauged subgroup
$G_{fc} \subset G$ ($G_{fc}$ can be as big as $G_{224}$ and as small as
$G_{213}$) to just the SM gauge symmetry $G_{213}$ [27] at the scale
$\Lambda_M$, while still preserving SUSY.

As a comment on the second assumption, it is of course possible to
construct a few MC singlet SUSY-preserving condensates in the most
attractive channels which would induce such a breaking pattern.  In
particular, in the preonic approach, one can construct a SUSY-preserving
condensate $\langle\Delta_R\rangle$, transforming as $(1, 3_R, 10^{*c})$
of $G_{224}$.  One assumes that this condensate $\langle\Delta_R\rangle$
in fact forms [6].  If it forms, it would naturally
be expected to be of order $\Lambda_M$ since it
conserves SUSY.
As is well-known, such a condensate breaks $G_{224}$ to
$G_{213}$.  It also breaks $L-R$ symmetry as well as $B-L$ so as to give
a heavy Majorana mass to composite $\nu_R$'s of order
$\Lambda_M$.  But such a pattern of breaking (i.e., $G \to G_{213}$),
which we assume, is
by no means unique and clearly needs justification.

Indeed, it seems to be a major challenge in any approach to higher
unification involving SUSY to know {\it why the preservation of the SM
gauge symmetry and that of SUSY} seem to go together.  And why are they {\it
both} stable, at least relative to Planck scale physics?  By the same
token, why {\it must} their breakings get correlated?  In the
conventional approach to grand unification with SUSY, this feature is
essentially put in by hand, to conform with observation, simply by a
choice of the Higgs multiplets and their parameters.  But even in the
superstring theories, at the present stage of our understanding, there
does not seem to be any {\it a priori} guarantee for such a correlation.
For instance, either SUSY or the SM gauge symmetry or both might have
broken already at the string unification scale.  But it is assumed in
the interest of progress, and rightly so, that the string vacuum somehow
picks out from among many classically allowed solutions the four-dimensional
space-time together with unbroken SUSY and the SM symmetry.
We will proceed by assuming that in the preonic approach as well there
exists such a correlation between the preservation of
SUSY and that of the SM symmetry.

One further comment is in order.  The breaking pattern assumed above
(e.g., $G_{224}\to G_{213}$) presumes that a dynamical breaking
of parity as well as of certain vectorial
symmetries like ``isospin'' and preon number $U(1)_V$
would be permissible at least in some SUSY QCD theories.  The
breakdown of such symmetries is, of course, not permitted in ordinary
QCD due to the no-go theorem of Vafa and Witten [28].  The proof of this
theorem does not, however, apply to massless SUSY QCD both because of
the presence of the gauge Yukawa interactions and also because of
masslessness of the matter fields.  It is tempting to conjecture
that massless SUSY QCD (with
$N_F$ and $N$ as listed above) in fact favors a dynamical breaking of
parity as well as of ``isospin'' and $U(1)_V$.  If true, there would
be a compelling reason why nature violates these symmetries [29].
The proof of this
conjecture remains a major challenge for the preonic approach and
is part of the same task
mentioned before -- i.e., to show the stability of the breaking pattern:
$\left( G\times (SUSY) \supset G_{224}\times SUSY\right)\to
G_{213}\times (SUSY)$.
\smallskip
\noindent
{\bf (C)~~Advantages of Combining the Ideas of Preons and Supersymmetry}:~~
I now list the advantages of the
preonic approach.  In addition to utmost economy in building blocks and
parameters, as described above, they include:
\smallskip
(1)~~{\bf Protection of the Masses of Quarks and Leptons}:  Utilizing
the Witten index theorem [30], which would forbid a dynamical breaking
of SUSY in the class of theories under consideration, except for the
presence of gravity, it is argued that chiral symmetry breaking preonic
condensates like $\langle\bar\psi\psi\rangle$, as well as the
metagaugino condensate
$\langle\lambda\cdot\lambda\rangle$, both of which break SUSY, are
necessarily damped by the factor $(\Lambda_M/M_{Pl})$, i.e.,
$$
\eqalignno{\langle \bar\psi^a\psi^a\rangle
&= a_{\psi_a}\Lambda_M^3(\Lambda_M/M_{Pl})\cr
\langle\lambda\cdot\lambda\rangle
&= a_\lambda\Lambda_M^3(\Lambda_M/M_{Pl})&(1)\cr}
$$
where $\Lambda_M$
denotes the scale parameter of the preonic metacolor force and $M_{Pl}$
denotes the Planck mass [7], and $a_{\psi_a}$ and $a_\lambda$ are
effective parameters of order unity.  With a perturbative input value for the
metacolor coupling near the Planck scale, the metacolor scale
$(\Lambda_M)$ is naturally small compared to $M_{Planck}$.  As a result,
the masses of $W$ and $Z$ as well as those of the composite quarks and
leptons are strongly damped compared to $\Lambda_M$ by the factor
$(\Lambda_M/M_{Pl})$.  This explains why quarks and leptons are so light
compared to their compositeness scale and thereby helps overcome the
first major hurdle of composite models.  In fact one obtains [6,7]:
$m_W\sim m_t\sim(1/10)\Lambda_M(\Lambda_M/M_{Pl})\sim 100~GeV$, where
$\Lambda_M$ is determined on independent grounds to lie around
$10^{11}~GeV$.  The anomaly-matching condition of 't Hooft is satisfied
trivially in our case because the unbroken symmetry is just
$G_{213}$ with the spectrum of $G_{224}$ which is, however, anomaly-free.
So the
relevant anomalies vanish at the preon and the composite levels.
\smallskip
{\bf (2)~~ Family Replication}: \ One can argue plausibly that the
composite
operator giving massless spin-1/2 quark (or lepton) consists of a minimum
of
three constituents [31], two of which come from flavor and
color-carrying preonic superfields and the third from the
metagluon vector superfield $(v_\mu ,\lambda ,\bar\lambda)$.
Recognizing that in a SUSY theory, fermionic
constituents can be interchanged by their boson partners (i.e. $\psi
\leftrightarrow \varphi$ and $v_{\mu} \leftrightarrow \lambda$ or
$\overline{\lambda}$ etc.), there exist several alternative three-particle
combinations with identical quantum numbers, which can make a left-chiral
$SU(2)_L$-doublet family $q^i_L$ -- e.g. (i)~$\sigma_{\mu \nu} \psi^f_L
\varphi^{c^*}_R v_{\mu \nu}$, (ii)~$\sigma^{\mu \nu} \varphi^f_L
\psi^{c^*}_R
v_{\mu \nu}$, (iii)~$\psi^f_L \psi^{c^*}_R \lambda$ and (iv)~$\varphi^f_L
(\sigma^{\mu} \overline{\lambda}) \partial_{\mu} \varphi^{c^*}_R$.  Here
$f=x$
or $y$ corresponds to up or down flavors and $c=(r,y,b)$ or $\ell$
corresponds
to the four colors.  {\it The plurality of these combinations, which stem
because of SUSY, is in essence the origin of family-replication}.  As
mentioned
before, by constructing composite superfields, Babu, Stremnitzer and I
showed [8] that at the level of minimum dimensional composite
operators
(somewhat analogous to $qqq$ for QCD) there are just three linearly
independent
chiral families $q^i_{L,R}$, and, in addition, two {\it vector-like
families}
$Q_{L,R}$ and $Q^{\prime}_{L,R}$, which couple vectorially to $W_L$'s and
$W_R$'s respectively.  Each of these composite families is, of course,
accompanied by its scalar superpartners.  To sum up, we see that owing to
a fermion-boson partnership in SUSY, the model provides a compelling reason for
replication and at least some rationale, subject to the assumption of
saturation
at the level of minimum dimensional composite operators, as to why the
number of
chiral families is three.  Clearly this last assumption needs further
justification.  Pending such a justification, however, let us still proceed
to
examine the masses of the fermions belonging to this system of {\it five
families} -- three chiral and two vector-like.  We will see that such a
system has some unique desirable features.
\smallskip
\noindent
(3)~~{\bf Inter-family Mass-Hierarchy}: ~~ First note that for the
purposes of
quantum numbers, the chiral and vector-like families can be represented by
$q_L
\sim$ ``$\psi^f_L \varphi^{c^*}_R v$'', $q_R \sim$ ``$\psi^f_R
\varphi_L^{c^*}
v$'', $Q_L \sim$ ``$\psi^f_L \varphi_L^{c^*} v$'', $Q_R \sim$
``$\varphi^f_L
\psi_L^{c^*} v$'', $Q_L^{\prime} \sim$ ``$\varphi^f_R \psi_R^{c^*} v$'' and
$Q_R^{\prime} \sim$ ``$\psi^f_R  \varphi_R^{c^*} v$''.  Utilizing these
compositions, one can see that the vector-families $Q_{L,R}$ and
$Q_{L,R}^{\prime}$ acquire relatively heavy masses through the metagaugino
condensate $<\vec{\lambda}\cdot\vec{\lambda}>$ of order $a_{\lambda}
\Lambda_M
(\Lambda_M/ M_{Pl}) \sim 1$ TeV, which are otherwise protected by the
$U(1)_X$
quantum number of the SUSY theory.  But the direct mass-terms of the
three chiral families $m^{(o)}
(q_L^i \rightarrow q_R^i)$ as well as the $Q-Q^{\prime}$ mixing terms
cannot be
induced through either $<\vec{\lambda}\cdot\vec{\lambda}>$ or
$<\bar{\psi}\psi>$.  These receive small contributions at most of order
(1/10-1)
MeV from products of  $<\bar{\psi}\psi>$ and $<\varphi^* \varphi>$
condensates,
each of which is damped by $(\Lambda_M/ M_{Pl})$.
The chiral families $q_{L,R}^i$ acquire masses primarily through their
mixings
with the vector-like families $Q_{L,R}$ and $Q_{L,R}^{\prime}$ which are
induced
by $<\bar{\psi}^a\psi^a>$.
Thus, dropping the
direct
and $Q-Q^{\prime}$ mixing mass terms, and ignoring QCD corrections for the
quarks for a moment, the Dirac-mass matrices of the five families for all
four
sectors - i.e., $q_u, q_d, l$ and $\nu$ - have the form [6,9]:
$$M_{f,c}^{(o)} =
\bordermatrix{& q_L^i & Q_L & Q_L^{\prime}\cr
\bar{q}_R^i & O & X\kappa_f & Y\kappa_c\cr
\bar{Q}_R & Y^{\prime\dagger}\kappa_c & \kappa_{\lambda} & O\cr
\bar{Q}_R^{\prime} & X^{\prime\dagger}\kappa_f & O &
\kappa_{\lambda}\cr} \eqno(2)$$
The index $i$ runs over three families.  The entities $X, Y, X^{\prime}$
and
$Y^{\prime}$ are three-component column matrices in the chiral
family-space having entries which
are
a of order unity.  In the above, $\kappa_f \equiv {\cal
O}(a_{\psi_f})\Lambda_M(\Lambda_M/ M_{Pl}), \kappa_c \equiv {\cal
O}(a_{\psi_c})\Lambda_M(\Lambda_M/ M_{Pl})$ and $\kappa_{\lambda} \equiv
{\cal O}(a_{\lambda})\Lambda_M(\Lambda_M/ M_{Pl})$. Since $\psi$'s are
in fundamental and $\lambda$'s in adjoint representation, we expect
$\kappa_\lambda$ somewhat larger than $\kappa_{f,c}$ -- i.e.,
$\kappa_{\lambda}\sim 1~TeV \approx (3-10)\kappa_{f,c}$.  As a
result,
the Dirac mass-matrices of all four sectors have {\it a natural see-saw
structure}.

In the absence of electroweak and QCD corrections,
due to left-right symmetry and full flavor-color independence of the metacolor
force, {\it not only $X^T = (X^{\prime})^T$ and $Y^T = (Y^{\prime})^T$, but
the
same $X, Y$ and $\kappa_{\lambda}$ apply to all four sectors: up, down,
charged leptons and neutrinos}.  Furthermore, ignoring electroweak
corrections (of order 5-10\%)
for a moment, one can always rotate the chiral quark or lepton fields
$q_R^i$ and $q_L^i$ to bring the row matrices $Y^T = (Y^{\prime})^T$ to the
simple form $(0,0,1)$ and simultaneously $X^T = (X^{\prime})^T$ to the form
$(0,p,1)$, with redefined $\kappa_f$ and $\kappa_c$.  Note the consequent
great
reduction even in effective parameters.
Upon examining the relevant preon-diagrams, one can argue that
the
effective parameter $p$ is less but not very much smaller than one.  A
value of
$p \simeq$ 1/3 to 1/5 is found to be quite natural.

Now, it may be seen that although the three chiral families are on par
as regards their binding and preon-content and thus the three entries in the
$X$ and likewise in the $Y$-matrices in the unrotated form (1) are expected to
be
comparable to each other, {\it the rank of the $5\times 5$ mass-matrix,
given by (1), automatically guarantees one linear combination ,
made primarily of the
three chiral families, to be massless, barring corrections of order
$({1\over 10}-1)~MeV$, which may be identified with the electron family
[9].}  At the same time the
heaviest chiral fermion (top) acquires a mass (in the above gross picture
)
of order 100-160 GeV.  In fact, we obtain (suppressing EW and QCD
corrections):
$$
\eqalignno{m_{u,d,e,\tilde\nu_e}^{(0)}
&\simeq 0+{\cal O}~({1\over 10}-1~MeV)\cr
m_{c,s,\mu ,\tilde\nu_\mu}^{(0)}
&\simeq (\kappa_f\kappa_c/\kappa_\lambda)(p^2/2)\cr
m_{t,b,\tau ,\tilde\nu_\tau}^{(0)}
&\simeq 2(\kappa_f\kappa_c/\kappa_\lambda)&(3)\cr}
$$

Here, $m_{\tilde\nu_i}^{(0)}$ denote the Dirac masses of the three
neutrinos, which, combined with the very heavy Majorana masses
$(\sim\Lambda_M)$ of the right-handed neutrinos gives the very light
$(\ltorder (1-50)eV)$ left-handed neutrinos.
For the $\mu -\tau$ mass-ratios, with universal $p$, we thus obtain:
$[m_{c,s,\mu}^{(0)}/m_{t,b,\tau}^{(0)}]\approx p^2/4$.  For $p\approx
1/3$ to $1/5$, which is not too small and natural, one obtains a large
$\mu -\tau$ hierarchy of $1/36$ to $1/100$, as observed.
In this way, one obtains {\it a very simple reason} for the inter-family
mass-hierarchy -- i.e., why $m_{u,d,e}\ll m_{c,s,\mu}\ll m_{t,b\tau}$.
In particular, one understands the gross pattern why $m_e \sim 1~MeV$,
while $m_t \sim
100-160~GeV$, so that $(m_e/m_t)\sim 10^{-5}$.  This is a major
achievement of the preonic approach.

In general, the effective parameter $p$ for quarks can differ from that
for leptons at $\Lambda_M$, by as much as perhaps a factor of 2, especially
if $SU(3)^c$ rather than $SU(4)^c$ is the gauge symmetry near the
Planck scale.  Unless electroweak corrections are substantially larger
than (the expected) 5-10\%, larger values of $m_t\approx 160-180~GeV$
would suggest $p(quark)\approx 1/4$ to $1/5$, while $p(leptons)\approx
1/3$.  The preon model can not,
of course, predict precise numbers including electroweak and QCD
corrections at $\Lambda_M$, because they depend upon strong
interaction matrix elements of preonic operators.  But it does
explain the {\it gross features}
of the inter-family mass-hierarchy, including the small number $(m_e/m_t)$,
based entirely on symmetries of
the theory, and thereby has removed the biggest surprises of the fermion
mass matrix.  The simplicity of the explanation lends support not only
for the preonic approach but also for the existence of the two
vector-like families, with masses of order $200~GeV - 1~TeV$, which are
crucial to the explanation of the inter-family mass-hierarchy.
Fortunately, these two families, especially their quark-members,
can be probed at the LHC and (the
leptonic members) at the NLC.

It turns out that by including electroweak corrections which break
left-right and up-down symmetries (so that $X\not= X^\prime$ and $Y\not=
Y^\prime$) as well as the direct mass-terms of the three chiral families
in the top-left $3\times 3$ block of (2), which are of order 1 MeV, one
also obtains a desirable pattern for CKM mixings and the deviation from
the universal $\mu -\tau$ hierarchy ratio of $p^2/4$, as desired [9].
\smallskip
\noindent
(4)~~{\bf A Unification of Scales:}~~ It is shown [6] that
the model is
capable of generating all the diverse scales -- from $M_{Planck}$ to
$m_{\nu}$
-- and thereby the small numbers such as $(m_W/M_{pl}) \sim (m_t/M_{pl})
\sim
10^{-17}, (m_C/M_{pl}) \sim 10^{-19}, (m_e/M_{pl}) \sim 10^{-22}$, and
$(m_{\nu}/M_{pl}) < 10^{-27}$ -- {\it in terms of just one fundamental
input
parameter}: the coupling constant $\alpha_M$ associated with the metacolor
force.  Corresponding to an input value $\overline{\alpha}_M \approx 1/27$
to
1/32 at $M_{Pl}/10$, the metacolor force generated by $SU(N)_M$ becomes
strong
at a scale $\Lambda_M \approx 10^{11} GeV$ for $N$=5 to 6.  Thus the first
big
step in the hierarchical ladder leading to the small number
$(\Lambda_M/M_{Pl})
\sim 10^{-8}$ arises naturally through renormalization group equations due
to
the slow logarithmic growth of $\overline{\alpha}_M$ and its perturbative
input
value at $M_{Pl}/10$.

The next step arises due to the constraint on SUSY breaking, which is
forbidden [30], except for the presence of gravity.  As mentioned
before, SUSY-breaking condensates like $\langle\lambda\lambda\rangle$
and $\langle\bar\psi \psi\rangle$ are thus naturally damped by
$(\Lambda_M/M_{Pl})[7]$.  These induce (a) SUSY-breaking mass-splittings
$\delta m_S\sim {\cal O}(\Lambda_M(\Lambda_M/M_{Pl}))\sim{\cal O}(1~TeV)$
and (b) $m_W\sim m_t\sim (1/10){\cal O}(\Lambda_M(\Lambda_M/M_{Pl}))$
$\sim {\cal O}(100~GeV)$.  Note the natural origin of the small numbers:
$(\delta m_s/M_{Pl})\sim 10^{-16}$ and $(m_W/M_{Pl})\sim 10^{-17}$.  As
also noted above, symmetries of the $5\times 5$ fermion mass-matrix take
us down to still lower scales -- in particular to $m_e\sim {\cal O}
(1~MeV)$, thus accounting for the tiny number $(m_e/M_{Pl}) \sim
10^{-22}$.

Finally, the familiar see-saw mechanism for neutrinos with
$m(\nu_R^i)\sim\Lambda_M\sim 10^{11}~GeV$ and $m(\nu^i)_{Dirac}\propto
\Lambda_M(\Lambda_M/M_{Pl})$ yields $m(\nu_L^i)\leq
10^{-3}~M_{Pl}(\Lambda_M/M_{Pl})^3\sim 10^{-27}M_{Pl}$.  In this way,
the model provides a {\it common origin} of all the diverse mass scales
-- from $M_{Pl}$ to $m_\nu$, and of the associated small numbers, as
desired [6].  This constitutes a unification of scales which is
fundamentally as important as the unification of forces.
\medskip
\noindent
(5)~~{\bf CP Violation:}~~  The model provides an elegant
mechanism for
spontaneous CP violation, which is shown to vanish (for the observed
processes),
if the masses of the electron family were set to zero [9], -- i.e. if
the
direct mass terms of the chiral families $m_{(0)}(q_L^i \rightarrow q^j_R)$
in
(1) are put strictly to zero.  Allowing for these direct mass-terms the
model
predicts $|\epsilon| \sim (m_d/m_b) \sin \xi \sim 2 \times 10^{-3}$ for a
maximal CP violating phase.  It also predicts an electric dipole moment for
the
neutron $\approx$ (1 to ${1\over20}) \times 10^{-25} ecm$, which is
observable.
\medskip
\noindent
(6)~~{\bf A Grand Fiesta of New Physics at {\bmit $10^{11}~GeV$}:}~~
The preonic approach identifies the
$10^{11}~GeV$ scale as its scale parameter $\Lambda_M$.  That
$\Lambda_M$ ought to be of order $10^{11}~GeV$ is in fact determined
within the model on two independent grounds.  First, because, in the
model, one obtains $m_W\sim \left({1\over
10}\right)~\Lambda_M(\Lambda_M/M_{Pl})$.  Equating this to 80 GeV, one
obtains $\Lambda_M\sim 10^{11}~GeV$.  Second, it turns out that a
meeting of the coupling constants is possible provided $\Lambda_M\sim
10^{11}~GeV$ [24].  A variety of phenomena thus occur at the
scale $10^{11}~GeV$ (see Fig. 2):~~
(a) First, the
metacolor force becomes strong at this scale; (b) thereby, the
composite quarks and
leptons as well as the composite Higgses -- i.e, the condensates -- are born;
(c)
the condensates in turn break certain
gauge and global symmetries of the model at the
scale $\Lambda_M$ while preserving SUSY (this includes $G_{224}\to
G_{213})$;
(d) in the process, the condensates also break spontaneously certain
fundamental
symmetries at the scale $\Lambda_M$ which include (i) parity, (ii) CP,
(iii) B-L and, as it turns out,
(iv) the Peccei-Quinn symmetry as well [32]; and last but not least,
(e) subject to a damping by the factor $(\Lambda_M/M_{Pl})$, the
condensates also break SUSY as well as the electroweak symmetry, thereby
giving desired masses $(\ltorder 1~TeV)$ to the missing SUSY partners,
W and Z and the quarks and leptons.  As mentioned before, the breaking
of $B-L$ occurs in a way that gives heavy Majorana masses $\sim
\Lambda_M$ to the composite $\nu_R$'s.
Finally, the preonic approach must also identify $10^{11}~GeV$ as
the {\it scale for inflation}, and $B-L$ violating baryo or
leptogenesis, because $10^{11}~GeV$
is the only scale available within the model, below Planck scale,
which could be relevant to the
necessary phase transition [33].

We see here that there is an enormous {\it economy of tools}:  One and
the same tool -- SUSY metacolor force coupled with gravity -- generates
dynamically all these phenomena.

It is furthermore interesting to note that, having determined
$\Lambda_M$ on two independent grounds {\it the preon model in fact
predicts that the PQ symmetry breaking scale as well as the Majorana
mass of the $\nu_R$'s are necessarily both of order $10^{11}~GeV$}.  It is
remarkable that both (i) the see-saw pattern for light neutrino masses
that is relevant to the MSW explanation for the solar neutrino puzzle,
together with the candidacy of $\nu^\tau$ for hot dark matter, and
independently (ii) astrophysical constraints on the axion restrict the
relevant scale in each case also to about $10^{11} - 10^{12}~GeV$.  In
summary, the preonic approach provides a genuine motivation for the
$10^{11}~GeV$-scale
and synthesizes a lot of new physics in terms of this single scale.  We
regard this feature an advantage of the preonic approach.
\medskip
\noindent
{\bf (7)~~Some Consequences and A Crucial Prediction:}
\item{(i)} $m_t\approx (100-180)~GeV$ [34].
\item{(ii)} New
contributions to $K^0\leftrightarrow \bar K^0$ and $K_L\to \bar\mu e$
from box graphs involving quarks and $W$'s as well as tree-level
$Z$-exchanges are smaller than the corresponding contributions from the
SM by 1 to 2 orders of magnitude, while those for $B^0-\bar B^0$ are
comparable to those of the SM.  Box diagrams containing squarks and
gluinos would be safe, as in all SUSY theories, {\it if} squarks of
different generations are sufficiently degenerate (e.g., to better than
5\% if $m_{\tilde q}\sim {\cal O}(3~TeV)$.  Such a degeneracy could arise
depending upon relative values of certain matrix elements of
preon-operators, but it is not {\it a priori} guaranteed by the model.
\item{(iii)}
$A(``Z''\to t\bar c)=A(t\to Zc)\approx \left({g_2\over \cos
\theta_W}\right)\left({\kappa_u\over\kappa_\lambda}\right)^2\left({p\over
2}\right)$$\approx \left({g_2\over\cos\theta_W}\right)\left(1/2\cdot 5
{}~{\rm to}~ 1/10\right)\%$, with $(\kappa_u/\kappa_\lambda)\approx (1/10~{\rm
to}~ 1/5)$
and $p\approx 1/5$.
\item{(iv)}
$\Delta m(D-\bar D)_{preon}\approx (1/2
-2)\times 10^{-14}~GeV$, while $\Delta m_{(expt)}\ltorder 1\cdot 3\times
10^{-13}~GeV$ and $\Delta m(SM)\approx (1-4)\times 10^{-15}~GeV$ [9].
\item{(v)}
$ A(Z\to \bar\mu e)$ which gives $B(\mu\to 3e)\approx (1 to 5)\times
10^{-13}$ [9].
\item{(vi)}
$\tau_\tau =\tau_\tau (SM)(1+\epsilon)=\tau_\tau
(SM)[1+(.01 ~{\rm to}~ .04]$, while $N_\nu (LEP)=3-2\epsilon=3-(.02
{}~{\rm to}~ .08)$,
where $\epsilon \equiv (\kappa_u/\kappa_\lambda)^2\approx (1/10-1/5)^2$.
 Compare with $N_\nu (LEP)_{expt}=2.988\pm .023$.  Note that the
corrections to $\tau_\tau$ and $N_\nu$ are correlated [35].
\item{(vii)}
prominent $\nu_\mu-\nu_\tau$ oscillations with $m(\nu_\tau)\approx
1~{\rm to}~
50~eV,~m(\nu_\mu)\approx (10^{-3}-10^{-1})eV$, $m(\nu_e)\approx
10^{-8}eV$ and $\nu_\mu-\nu_e$ mixing consistent with MSW solution [36].
\item{(viii)} $(edm)_n\approx (10-1/2)\times 10^{-26}ecm$ [9].
\item{(ix)} Existence
of SUSY partners with masses $\approx 100~GeV-2~TeV$ and two Higgs
doublets corresponding to $\tan \beta\approx 30-40$.
\item{(x)}
\undertext{A Crucial Prediction and Hallmark of the Model}:
There must exist
precisely two {\it complete} vector-like quark-lepton families (no more,
no less) with masses:  $m(E,E^\prime ,N,N^\prime)\approx
(150-700)GeV; m(U,D,U^\prime ,D^\prime)\approx(350-1700)~GeV$ [6-9, 37].
{\it
They are not only predicted by the model, but they serve an
essential purpose in generating masses for the observed quarks and
leptons and also explaining their inter-family mass-hierarchy.  }
Furthermore, their existence is fully compatible with measurements of
the $\rho$ and the $S$ (or $\epsilon_3$) parameters, because their
masses are $SU(2)_L\times U(1)$-symmetric.

For these reasons, I believe that it is a {\it good bet} that precisely
two vector-like families -- especially the quark members -- await
discovery at the LHC and certainly at a future version of the SSC.
Likewise, the leptonic members $(N,N^\prime ,E,E^\prime)$ await
discovery at the next linear $e^-e^+$ collider $(E_{cM}\approx
500~GeV-1~TeV)$.  If light enough, they can even be produced singly at
LEP 200 [37]:  $e^-e^+\to N+\bar\nu_\tau\to (Z+\nu_\tau)+\bar\nu_\tau\to
(e^-e^+)_Z+\nu_\tau +\bar\nu_\tau$.
\medskip
\noindent (8.)~~{\bf Supersymmetry Breaking:}
The preonic theory requires the presence of a new metacolor force
which becomes strong at a superheavy scale $\Lambda_M >>
1$~{\it TeV}.   This force influences {\it directly} the observed sector
of quarks, leptons and Higgs bosons, since it operates on their
constituents -- the preons.  This is analogous to QCD influencing
observed hadrons.  The existence of such
a force, in the presence of
gravity,  allows the possibility of a dynamical breaking of supersymmetry
directly in the observable sector (albeit with a damping by the Planck
mass),
which thus transmits efficiently into the masses of the squarks, the
gluinos and
the winos.  This may well be an advantage over a dynamical breaking of
supersymmetry occurring entirely in the hidden sector of a superstring
theory,
which seems to be the only possibility in such a theory if quarks and
leptons are elementary.  For a preonic theory on the other hand,
assuming that it arises from a superstring theory, there is the general
possibility that SUSY-breaking has its roots in both the observable and
the hidden sectors and the two breakings influence each other.
Alternatively, the breaking in one sector may drive that in the other.
Even though SUSY-breaking effects induced through the hidden sector are
expected to be small compared to $M_{Pl}$ (and perhaps also
$\Lambda_M$), they may still have significant effects in removing
certain degeneracies (flat directions) and thereby influencing the
dynamics in the observable sector (see remarks at the end of Ref. 26).
\bigskip
\noindent
{\rmb 4.~~ Unity of Forces at the Preon Level}

Although the derivation of a preon model resembling that proposed in
Ref.~6 from
a superstring theory is still awaited, it is intriguing to ask whether the
coupling constants $g_1\,, g_2$ and $g_3$ extrapolated in the context of
the
preon model from their measured values at low energies do indeed meet with
each
other as well as with $g_M$ near the Planck scale, for any reasonable
choice of
the metacolor gauge symmetry $G_M\,$ and the flavor-color gauge symmetry
$G_{fc}$ which operates near the Planck scale.  The precise nature of $G_M$
and
$G_{fc}$ may hopefully get determined ultimately by an underlying
superstring
theory, if preons have their origin from such a theory.  In this case,
despite the
non-unifying appearance of the effective symmetry $G_M \times G_{fc}$, the
constraints of grand unification including the quantization of electric
charge
and the familiar equality of the coupling constants near the unification
scale
$M_{SU}$ would still hold, especially for $k=1$ Kac-Moody algebra, barring,
of
course, Planck-scale threshold effects.

Note that the extrapolation based on renormalization group equations is
fully
determined in regions I and II(see Fig. 3)  because the spectrum and the
gauge
symmetry are fixed, while in region III, there is only a few discrete
choices
which can be made as regards the metacolor gauge symmetry $G_M$ and the
flavor-color gauge symmetry $G_{fc}$.  The scale $\Lambda_M$ is fixed at
about
$10^{11}$ GeV (within a factor of 3, say) by requiring consistency with the
hierarchy of scales, in particular by the observed mass of $m_W$.

Babu, Parida and I [24,38] found that with the measured
low-energy
values of $g_1, g_2$ and $g_3$ at LEP and thus with $\sin^2\theta_W \simeq
.2333,
\alpha = 1/127.9$ and $\alpha_3 = .118$ at $m_Z$ as inputs, the coupling
constants including $g_M$, show a clear tendency to converge to a common
value
within a few percent of each other for $G_M=SU(5)_M$ [24], as well
as $G_M=SU(6)_M$ (with
the inclusion of threshold effects at $\Lambda_M$ [38]) at a scale
$M_U \approx (2-5)
\times
10^{18}$ GeV, for example for the choice $G_{fc} =
SU(2)_L
\times U(1)_{I_{3R}} \times SU(4)^c$, (see Fig. 2).  Such a convergence,
which is clearly non-trivial, demonstrates
that the unity of forces may well occur at the  level of preons in a manner
that
is truly novel compared to the conventional approach of elementary quarks
and
leptons.
\bigskip
\noindent
{\rmb 5.~~Summary and Concluding Remarks}

The preonic approach which has evolved in isolation over the last few
years [6-9,24] exhibits some very desirable properties:  (i) Utilizing
local supersymmetry to its advantage, it seems capable of addressing the
issues of the origin of families, and in particular that of inter-family
mass-hierarchy and diverse mass-scales.  (ii) It provides the scope for
synthesizing a rich variety of phenomena as having a {\it common origin}
at $\Lambda_M$ (see Fig. 2).  (iii) It is viable.  And, (iv) most
important, it is falsifiable.  As such I believe, despite its
unconventional status, that there is a good chance that it may even
possess a certain degree of truth and that it may well provide the
right effective theory to describe physics between the Planck scale and
$\Lambda_M$.  It should still (and I believe it must if
it is the right theory) have its origin within a superstring theory,
which should provide a good quantum theory of gravity and a unified
origin of all the forces including gravity.  The superstring theory, in
this case, would need to yield the preonic spectrum and its gauge
symmetry, but it would be relieved from the burden of yielding the
right package of Higgs-sector parameters, because there is no elementary
Higgs sector in the preonic approach.

The main disadvantage of the preonic approach at present is that it
rests on two dynamical assumptions.  As stated before, they are:
(i) The asymptotically free SUSY metacolor force (with $N_F=N, N+1$, or
$N+2$ and $m=0$) confines and (ii) the metacolor force, combined with
the (weaker) flavor-color forces, breaks the original global symmetry,
that contains the flavor-color gauge symmetry $G_{fc}$ (which could be
either $G_{224},~G_{2213},~G_{214}$ or even $G_{213})$, into just the SM
symmetry $G_{213}$ at $\Lambda_M$.  This second assumption requires that
at least the relevant class of locally supersymmetric QCD theories
(unlike ordinary QCD) break vectorial symmetries such as isospin (this
induces $\kappa_d\not= \kappa_u$ at $\Lambda_M$) and preon number, as
well as parity (if $G_{fc}=G_{224}$).  It is a challenge to show that
these symmetries are indeed broken in certain locally supersymmetric QCD
theories.  A positive (or negative) result in this respect will strongly favor
(or disfavor)
the preonic approach.  Ascertaining the answer to this issue
is thus a major task for this approach.

On the positive side, while the pattern of condensates (e.g., the
formation of $\langle \Delta_R\rangle$) need to be assumed within the
preonic approach, the scales of the condensates including those with a
damping by $(\Lambda_M/M_{Pl})$ are motivated on general grounds.
Furthermore, even without being derived so far from a superstring
theory, {\it it is still the most economical model around}.  In
addition, it provides, as mentioned above, simple explanations for
certain major puzzles, which the conventional approach so far has not.
As such, I believe, it deserves far more attention as regards the study of
all its aspects, in particular those involving properties of locally
supersymmetric QCD theories, than it has received so far.  Furthermore,
its derivation from a superstring theory (including variants -- e.g., as
regards choice of metacolor and flavor-color gauge symmetries) is, of
course, most desirable.

An additional remark, if the preonic ideas presented here turn out to
be right (at least in essence), {\it one would clearly need an understanding
of the internal dynamics of highly relativistic composite systems
possessing confinement, for which the constituents as well as the
composites are essentially massless compared to the compositeness
scale.}
Our familiar ideas about bound states (not to mention potential models)
will clearly not apply.  And one would need a new SUSY QCD bag model, which may
be as novel relative to the idea of the QCD bag as the latter is relative to
our view of the
composite nuclei,bound by short-range forces without confinement,
and likewise the nuclei (bound by short range forces) are
relative to the atoms.  In short, based on past experiences, it is
not unreasonable to expect
novelty at every stage of compositeness.  The question is:  have we
exhausted the bag of such novelties, or, is there still more to come?

Before concluding, it is worth recalling that there have been occasions
where the idea of a certain symmetry has turned out to be right, yet
it did not
progress for quite a while because of its association with the wrong
fields.  In particular, the idea of
the Yang-Mills symmetry which was originally associated with the isospin
degree of freedom of the observed proton and neutron and that of the
SU(3)-flavor symmetry which in the beginning was associated with the
observed ($p,n$ and $\Lambda$) as triplet, are both examples of this
kind.  These ideas did not succeed until it was realized that they ought
to be associated with
the {\it constituents} of ($p,n$ and $\Lambda$) -- i.e., quarks with
color.  One may thus wonder whether the ideas of superstrings and grand
unification ought to be associated with the observed quarks and leptons,
or with a new layer of constituents -- the preons.  I believe that the
examples cited above call for some caution in the conventional view in
this regard and
warrants an {\it open point of view}.

To conclude, if there is something that I feel more certain about, in
addition to the existence of the conventional SUSY and the Higgs
particles, it is the existence of the two vector-like
families [6-9].  This is because of the compelling nature of their origin and
the simplicity with which they explain the inter-family mass hierarchy.
In spite of the promising features of the preonic approach, there is
of course a good chance that one may be fooling
oneself about its prospects.  {\it How can one really tell}?  In answer to
this question,  it is
comforting to know that the approach provides at least one crucial test.
{\it The existence or non-existence of the two vector-like families in the
mass-range of a few hundred GeV to about 2~TeV would vindicate or totally
falsify the preonic approach, as presented here}.  There are, of
course, other intriguing flavor-changing effects which could show,
at LEPI,
LEP II or Tevatron (see Sec. 3.7).
It is a great relief that the LHC has finally been approved and there is
good prospect for approval of the NLC in Japan.  These can search for
the Higgses, SUSY {\it and} the vector-like families.  It is these
experimental facilities which could ultimately free us from the
present bottleneck in particle physics
and hopefully tell us which of our preconceived notions about elementary
particles are right, if any, and which are wrong.
\bigskip
\noindent
{\rmb 6.~~Acknowledgements}

The research described in this talk is supported in part by a grant from
the U.S. National Science Foundation and in part by a sabbatical-leave
grant from the University of Maryland.  This talk was prepared during the
author's visit to the Yukawa Institute of Theoretical Physics, Kyoto,
Japan, which was supported by a generous grant by the Government of
Japan.  The hospitality of Kenzo Inoue and Takeo Inami at the Yukawa
Institute and that of
Edward Witten and Frank Wilczek at the Institute for Advanced Study,
where this manuscript was prepared,
is gratefully acknowledged.  It is a
pleasure to thank Anthony Sanda and S. Suzuki for arranging an excellent
meeting at Nagoya and for their hospitality.  I would also like to thank
Valerie Nowak for the most efficient typing of this manuscript.
\bigskip
\centerline{\rmb References}
\item{[1]}
J.C. Pati and Abdus Salam; Proc. 15th High Energy Conference, Batavia,
reported by J.D. Bjorken, Vol. 2, p. 301 (1972); Phys. Rev. {\bf 8}
(1973) 1240; Phys. Rev. Lett. {\bf 31} (1973) 661; Phys. Rev. {\bf D10}
(1974) 275.
\item{[2]}
H. Georgi and S.L. Glashow, Phys. Rev. Lett. {\bf 32} (1974) 438.
\item{[3]}
H. Georgi H. Quinn and S. Weinberg, Phys. Rev. Lett. {\bf 33} (1974)
451.
\item{[4]}
J. Wess and B. Zumino, Nucl. Phys. {\bf B70} (1974) 39; Phys. Lett. {\bf
49B} (1974) 52; D. Volkov and V.P. Akulov, JETP Lett. {\bf 16} (1972)
438.
\item{[5]}
For a review see E. Farhi and L. Susskind, Phys. Rev. {\bf 74} (1981)
277 and references therein.
\item{[6]}
J.C. Pati, Phys. Lett. {\bf B228} (1989) 228.
\item{[7]}
J.C. Pati, M. Cvetic and H. Sharatchandra, Phys. Rev. Lett. {\bf 58}
(1987) 851.
\item{[8]}
K.S. Babu, J.C. Pati and H. Stremnitzer, Phys. Lett. {\bf B256} (1991)
206.
\item{[9]}
K.S. Babu, J.C. Pati and H. Stremnitzer, Phys. Rev. Lett. {\bf 67}
(1991) 1688.
\item{[10]}
M. Green and J. Schwarz, Phys. Lett. {\bf 149B} (1984) 117; D. Gross, J.
Harvey, E. Martinec and R. Rohm, Phys. Rev. Lett {\bf 54} (1985) 502; P.
Candelas, G. Horowitz, A. Strominger and E. Witten, Nucl. Phys. {\bf
B258} (1985) 46.
\item{[11]}
SO(10):  H. Georgi, Proc. AIP Conf. Williamsburg (1994); H. Fritzsch and
P. Minkowski, Ann. Phys. (NY) {\bf 93} (1975) 193.  E(6):  F. Gursey, P.
Ramond and P. Sikivie, Phys. Lett. {\bf 60B} (1976) 177.
\item{[12]}
J.C. Pati and A. Salam, Phys. Rev. {\bf D10} (1974) 275; R.N. Mohapatra
and J.C. Pati, Phys. Rev. {\bf D11} (1975) 566; Phys. Rev. {\bf D11}
(1975) 2558; G. Senjanovic and R.N. Mohapatra, Phys. Rev. {\bf
D12}(1975) 1502.
\item{[13]}
For example, a two-step breaking of SO(10) into the SM symmetry as well
as the preonic approach [6] lead to $M_{\nu_R}\sim 10^{11}-10^{12}~GeV$.
\item{[14]}
Particle Data Group, Review of Particle Properties, Phys. Rev. {\bf
D45}, Part II, SI-S584, June 1, 1992.
\item{[15]}
LEP data, Particle Data Group (June, 1994).
\item{[16]}
S. Dimopoulos and H. Georgi, Nucl. Phys. {\bf B193} (1981) 150; N.
Sakai, Z. Phys. {\bf C11} (1982) 153.
\item{[17]}
P. Langacker and M. Luo, Phys. Rev. {\bf D44} (1991) 817; U. Amaldi, W.
de Boer and H. Furstenau, Phys. Lett. {\bf B260} (1991) 447; J. Ellis,
S. Kelley and D.V. Nanopoulos, Phys. Lett. {\bf B260} (1991) 131; F.
Anselmo, L. Cifarelli, A. Peterman and A. Zichichi, Nuov. Cim. {\bf
A104} (1991) 1817.
\item{[18]}
P. Langacker, Review talk at Gatlinburg Conference, June '94,
HEP-PH-9411247.
\item{[19]}
For analysis of this type, see e.g. R. Arnowitt and P. Nath, Phys. Rev.
Lett. {\bf 69} (1992) 725; K. Inoue, M. Kawasaki, M. Yamaguchi and T.
Yanagida, Phys. Rev. {\bf D45} (1992) 328; G.G. Ross and R.G. Roberts,
Nucl. Phys. {\bf B377} (1992) 571; and J.L. Lopez, D. Nanopoulos and H.
Pois, Phys. Rev. Lett. {\bf 47} (1993) 2468.  Other references may be
found in the last paper.
\item{[20]}
J.C. Pati, A. Salam and U. Sarkar, Phys. Lett. {\bf 133B} (1983) 330;
J.C. Pati, Phys. Rev. Rap. Comm. {\bf D29} (1983) 1549.
\item{[21]}
H. Kawai, D. Lewellen and S. Tye, Nucl. Phys. {\bf B288} (1987) 1;
I. Antoniadis, C. Bachas and C. Kounnas, Nucl. Phys. {\bf B289} (1987)
187.
\item{[22]}
Some partially successful three-family solutions with top acquiring a
mass of the right value ($\approx 175~GeV$) and all the other fermions
being massless at the level of cubic Yukawa couplings have been obtained
with $Z_2\times Z_2$ orbifold compactification (A. Farragi, Phys. Lett.
{\bf B274} (1992) 47; Nucl. Phys. {\bf B416} (1994) 63; J. Lopez, D.
Nanopoulos and A. Zichichi, Texas A\&M preprint CTP AMU-06/95).  But
all the other masses and mixings including $m_e\sim {\cal O}(1~MeV)$ are
attributed to in-principle calculable higher dimensional operators.  It
seems optimistic that the entire package of effective parameters
would come out correctly this way with the desired hierarchy.  But, of
course, there is no argument that they cannot.  Thus it seems most
desirable to pursue this approach as far as one can.  This is why I keep
an open mind with regard to both the conventional approach and the
preonic alternative.
\item{[23]}
V.S. Kaplunovsky, Nucl. Phys. {\bf B307} (1988) 145; recently, K. Dienes
and A.E. Farragi (preprint IASSNS-HEP 94/113) provide general arguments
why string-threshold corrections arising from the massive tower of
states are naturally suppressed and,
thus, these corrections do not account for such a mismatch between
the two scales.
\item{[24]}
K.S. Babu and J.C. Pati, Phys. Rev. Rap. Comm. {\bf D48} (1993) R1921.
\item{[25]}
I. Affleck, M. Dine and N. Seiberg, Nucl. Phys. {\bf B241} (1984) 493;
Nucl. Phys. {\bf B256} (1985) 557; D. Amati, K. Konishi, Y. Meurice,
G.C. Rossi and G. Veneziano, Phys. Rep. {\bf 162} (1988) 169 and
references therein.
\item{[26]}
Using the power of holomorphy, N. Seiberg has recently studied extensively
the vacuum properties of a large class of strongly interacting SUSY
gauge theories [N. Seiberg, IASSNS-HEP-94/57, Pascos '94 talk and
references therein], including the cases of relevance to us -- i.e.,
$N_f=N,~N+1$ or $N+2$.  Allowing for (a) possible disconnected branches to go
near the origin of scalar field-space, (b) the presence of flavor-color
gauge interactions, and (c) the fact that a minimum of three-body (like
$\sigma_{\mu\nu}\psi\varphi^*v_{\mu\nu}$) rather than two-body
$(\psi\varphi^*)$ composites seem to be needed for dynamical
consistency of massless spin-1/2 composites made of massless spin-1/2
and spin-0 constituents [31,8], I have not yet been able to
disentangle whether the
symmetry breaking pattern $G\to G_{213}$, or rather
$G_{exact}=[(G_{fc})_{gauge}\times U(1)_V\times U(1)_X]\to G_{213}$
(where $G_{fc}$ may even be as small as just $G_{213}$), which I assume,
and also the associated spectrum of chiral {\it and} vector-like
families obtained in Ref. 8,
is compatible with Seiberg's analysis or not.  Note that the masses of
these chiral and vector-like families are protected by SUSY and the
massless spectrum satisfies the 't Hooft anomaly-matching condition
because the unbroken symmetry $G_{213}$ belongs to $G_{224}$, which
governs the spectrum, and is anomaly-free.  One last remark:  Soft
SUSY-breaking scalar preon $(mass)^2$-term of the form $m_0^2 \sum_a
(|\varphi_L^a|^2+|\varphi_R^a|^2)$ is in general expected to be induced
near the Planck scale through the superstring-generated hidden sector
dynamics (as in the conventional approach).  Considering that such
mass-terms remove the degeneracy and favor the solution near
$\varphi_{L,R}=0$, the question arises as to how they would affect the
Seiberg analysis, which is based on extrapolating from
$\varphi_{L,R}=\propto$ to
zero.  In spite of such mass-terms, one of the main features of the
preonic approach, that rests on the damping of
$\langle\bar\psi\psi\rangle$ and $\langle\lambda\lambda\rangle$ by the
Planck mass (see text) should, however, still hold, especially if $m_0
\ltorder \Lambda_M~(\Lambda_m/M_{Pl})$.  These questions are under
investigation.
\item{[27]}
In general, some discrete subgroups of $U(1)_V$ and $U(1)_X$ may
survive.
\item{[28]}
C. Vafa and E. Witten, Nucl. Phys. {\bf B234} (1984) 173; Phys. Rev.
Lett. {\bf 53} (1984) 535.
\item{[29]}
It is interesting that Seiberg finds violation of $U(1)_V$ in certain
cases of SUSY-QCD theories (see Ref. 26).
\item{[30]}
E. Witten, Nucl. Phys. {\bf B185} (1981) 513; {\bf B202} (1983) 253; E.
Cohen and L. Gomez, Phys. Rev. Lett. {\bf 52} (1984) 237.
\item{[31]}
J.C. Pati, Phys. Lett. {\bf 144B} (1984) 375.
\item{[32]}
K. Babu, K. Choi, J.C. Pati, X.M. Zhang, Phys. Lett. {\bf B333} (1994)
364.
\item{[33]}
For such a picture of inflation see M. Cvetic. T. H\"ubsch, J. Pati and
J. Stremnitzer, Phys. Rev. {\bf D40} (1989) 1311.
\item{[34]}
This allows for gauge and Yukawa radiative effects [J.B. Kim, K. Babu
and J. Pati (to appear)] and for the possibility that the
effective parameter $p$ for quarks could differ from that of leptons by
as much as a factor
of two at $\Lambda_M$ (see Sec. 3.7).  A rough upper limit of about 150
GeV for $m_t$ obtained in Ref. 9, was based on considerations which do
not include these effects.
\item{[35]}
K. Babu, J. Pati and X.M. Zhang, Phys. Rev. {\bf 46} (1992) 2190.
\item{[36]}
K. Babu, J. Pati and H. Stremnitzer, Phys. Lett. {\bf 264} (1991) 347.
\item{[37]}
K. Babu, J. Pati and H. Stremnitzer, UMD-PP-94-99, Phys. Rev. {\bf D}
March (1995).
\item{[38]}
K. Babu, M. Parida and J. Pati, (to appear).
\vfill\eject

\centerline{\rmb Figure 2}
\noindent
{\bf A grand fiesta of new physics at ${\bmit 10^{11}~GeV}$:}
The preonic approach suggests the existence of a rich variety of new
physics, listed above, at the $10^{11}~GeV$ scale, all of which
could have a {\it common origin} through one and the same source:  the locally
supersymmetric metacolor force, operating in the observable sector, with
a scale-parameter $\Lambda_M\sim 10^{11}~GeV$.  As the metacolor force
becomes strong at this scale, it generates the set of phenomena,depicted
above, some of
which preserve SUSY (such as the breakings of $SU(4)^c$, B-L and PQ
symmetry) and some which do not (such as $\delta m_s,~m_q$ and $m_W$).
Since SUSY-breaking effects {\it need} the collaboration of gravity
(even though perturbative) with the metacolor force, they are naturally
damped compared to the metacolor scale by the gravitational coupling.
Thus arises the hierarchy of scales -- e.g., $m_W\sim m_t\sim\delta
m_s\sim \Lambda_M(\Lambda_M/M_{Pl})\sim
M_{Pl}(\Lambda_M/M_{Pl})^2\ll M_{Pl}$.

\bye